\documentclass[twocolumn,showpacs,preprintnumbers,amsmath,amssymb]{revtex4}

\usepackage{graphicx}
\usepackage{dcolumn} 
\usepackage{bm}

\begin{document}

\title{Room-temperature ferromagnetism in Sr$_{1-x}$Y$_x$CoO$_{3-\delta}$ 
($0.2\leq x \leq 0.25$)
}

\author{W. Kobayashi}
\email{kobayashi-wataru@suou.waseda.jp}
\affiliation{Department of Applied Physics, Waseda University, 
Tokyo 169-8555, Japan}

\author{S. Ishiwata}
\affiliation{Department of Applied Physics, Waseda University, 
Tokyo 169-8555, Japan}

\author{I. Terasaki}
\affiliation{Department of Applied Physics, Waseda University, 
Tokyo 169-8555, Japan\\
also CREST, Japan Science and Technology Agency, Tokyo 108-0075, Japan}

\author{M. Takano}
\affiliation{Institute for Chemical Research, Kyoto University, 
Uji 611-0011, Japan}

\author{I. Grigoraviciute}
\affiliation{Materials and Structures Laboratory, Tokyo Institute of 
Technology, Yokohama 226-8503, Japan}

\author{H. Yamauchi}
\affiliation{Materials and Structures Laboratory, Tokyo Institute of 
Technology, Yokohama 226-8503, Japan}

\author{M. Karppinen}
\affiliation{Materials and Structures Laboratory, Tokyo Institute of 
Technology, Yokohama 226-8503, Japan}

\date{\today}

\begin{abstract}

We have measured magnetic susceptibility and resistivity 
of Sr$_{1-x}$Y$_x$CoO$_{3-\delta}$ ($x=$ 0.1, 0.15, 0.2, 0.215, 0.225, 0.25, 0.3, and 0.4), 
and found that Sr$_{1-x}$Y$_x$CoO$_{3-\delta}$ is a room temperature ferromagnet 
with a Curie temperature of 335 K in a narrow compositional 
range of 0.2 $\leq x\leq$ 0.25. This is the highest transition temperature among 
perovskite Co oxides. 
The saturation magnetization for $x=$ 0.225 is 0.25 $\mu_B$/Co at 10 K, 
which implies that the observed ferromagnetism is a bulk effect. 
We attribute this ferromagnetism to a peculiar Sr/Y ordering. 

\end{abstract}


\maketitle

\section{\label{sec:level1}Introduction}

Transition-metal oxides with perovskite-based structure exhibit 
many fascinating phenomena such as colossal magnetoresistance 
in Mn oxides \cite{Mn}, 
high temperature superconductivity in Cu oxides \cite{HTSC}, and 
ferroelectricity in Ti oxides \cite{Ti}. 
Perovskite Co oxides also show very rich physics 
such as spin-state transition \cite{Co1, Co2}, 
ferromagnetism \cite{LaCoO}, and thermoelectricity \cite{thermo}. 
The ferromagnetism in the perovskite Co oxide La$_{1-x}$Sr$_x$CoO$_3$ 
has been explained in terms of 
the double-exchange interaction between Co$^{3+}$ and Co$^{4+}$ 
in the intermediate spin state. 
The ferromagnetic transition temperature ($T_c$) is
270 K at highest, which is significantly lower than $T_c$=370 K of 
perovskite Mn oxides at an optimum composition. 

In recent years, the A-site ordered perovskite $R$BaCo$_2$O$_{5.5}$ 
($R=$ rare earth element) has been intensively studied \cite{taskin, LnBaCoO1, LnBaCoO2}. 
It shows a ferromagnetic transition around 300 K (higher than $T_c$ of La$_{1-x}$Sr$_x$CoO$_3$) 
with giant magnetoresistance \cite{taskin, MR}.
In addition, a transition from the ferromagnetic state to an antiferromagnetic state 
is seen around 255 K in the magnetization \cite{LnBaCoO2}. 
The complicated magnetism has been explained as follows:
The Co$^{3+}$ ions in the intermediate spin (IS) state of $S=1$
exhibit an orbital ordering at 350 K along the $a$ axis \cite{LnBaCoO1},
which induces ferromagnetic interaction along the $a$ axis.
On the other hand, antiferromagnetic interaction is expected 
along the $b$ axis because of the conventional superexchange, 
which causes the antiferromagnetic order below 270 K.
At higher temperatures, thermally excited carriers induce
the double-exchange interaction to align the ferromagnetic chains,
and stabilizes the ferromagnetic state from 270 to 300 K \cite{LnBaCoO1, taskin}.

Very recently, Withers {\it et al.} \cite{withers, james}
and Istomin {\it et al.} \cite{istomin1, istomin2} independently reported 
a new A-site ordered perovskite Co oxide Sr$_{1-x}R_{x}$CoO$_{3-\delta}$ 
($R$ = Y and lanthanide). 
According to their structure analysis, Sr and $R$ are ordered 
because of the different ionic radii.
The $R$ ions prefer to occupy at every two sites in the $ab$ plane, which doubles
the lattice parameter of the $a$ and $b$ axes from the primitive 
perovskite cell.
The ordered $ab$ planes are stacked along the $c$-axis 
in units of four cells, 
and the $c$ axis length quadruples that of the primitive cell. 
This will give an ideal composition of Sr$_{0.75}R_{0.25}$CoO$_{3-\delta}$,
but a particular composition near $x=$1/3 has been extensively studied so far.
Aside from the cation ordering, the two groups proposed different oxygen 
ordering structures, and the complete structure is still controversial.

Magnetic properties of Sr$_{1-x}$Y$_{x}$CoO$_{3-\delta}$
were measured only for $x$=0.1 and 0.33 \cite{goosens, maignan}, 
and we expected that the peculiar Sr/Y ordering near $x=1/4$
could show a novel magnetism. 
We have found that polycrystalline samples of 
Sr$_{1-x}$Y$_x$CoO$_{3-\delta}$ show a ferromagnetic transition at 335 K
in a narrow range of $0.2\leq x \leq 0.25$, 
which is the highest transition temperature among perovskite Co oxides. 
The magnetization of the $x=$ 0.225 sample is 0.25 $\mu_B$/Co at 10 K,
indicating a bulk ferromagnet. 
We suggest that this ferromagnetism is driven by the orbital ordering, 
just like the ferromagnetism in $R$BaCo$_2$O$_{5.5}$.
However, $T_c$ of the former is even higher than that of 
the latter, and ferromagnetic order survives down to 5 K. 
We will discuss a possible origin for the difference.

\section{Experiment}

Polycrystalline samples of Sr$_{1-x}$Y$_x$CoO$_{3-\delta}$ 
($x=$ 0.1, 0.15, 0.2, 0.215, 0.225, 0.25, 0.3 and 0.4) and Sr$_{0.775}R_{0.225}$CoO$_{3-\delta}$ 
($R=$Dy, Ho, and Er) were prepared by a solid-state reaction. 
Stoichiometric amounts of SrCO$_3$, Y$_2$O$_3$, Dy$_2$O$_3$, Ho$_2$O$_3$, Er$_2$O$_3$, 
and Co$_3$O$_4$ were mixed, and the mixture was calcined 
at 1100$^{\circ}$C for 12h in air. 
The product was finely ground, pressed into a pellet, 
and sintered at 1100$^{\circ}$C for 48h in air, followed by slow 
cooling down to room temperature in the furnace.
That the resultant oxygen content of the sample represented an 
equilibrium value (in air) was confirmed by means of thermogravimetric annealing experiments.
The oxygen content was determined through iodometric titration\cite{titration}.
A powdered sample of 20-40 mg was dissolved in 1 M HCl solution ($ca.$ $30$ ml) containing an excess 
of KI. The titration of the formed I$_2$ 
was performed against a standard 0.015 M sodium thiosulphate 
solution using starch as an indicator.
For each sample, the analysis was repeated three times in minimum 
with a reproducibility better than $\pm $0.003 for $\delta$.
Throughout this paper, we will use the nominal composition $x$,
although the third decimal point of $x$ includes some uncertainty.
We evaluated through energy dispersive X-ray analysis (EDX) 
the real compositions for $x=$ 0.215 and 0.225 to be 
Sr $:$ Y $=0.781:0.219$ and $0.766:0.234$, respectively,
meaning that $x$ is correct within an accuracy of $\pm$0.01.
As a reference, Sr$_{0.775}$Y$_{0.225}$CoO$_{3-\delta}$ 
powder was sealed with KClO$_4$ in a gold capsule, and then 
treated at 3 GPa and 700$^{\circ}$C for 30 min in 
a conventional cubic anvil-type high pressure apparatus. 


The X-ray diffraction (XRD) of the sample was measured using 
a standard diffractometer with Fe K$_\alpha$ radiation as an X-ray source 
in the $\theta$-$2\theta$ scan mode. 
The resistivity was measured by a four-probe method, 
below room temperature (4.2-300~K) in a liquid He cryostat, 
and above room temperature (300-400~K) in an electronic furnace. 
The magnetization $M$ was measured from 5 to 400 K 
by a commercial superconducting quantum interference device 
(SQUID, Quantum Design MPMS). We applied 0.01-0.1 T for the measurement of 
$M$ ($T$) data, and -7 to 7 T for $M$ ($H$) data. 

\section{Results and Discussion}

\begin{figure}
\vspace*{0cm}
\includegraphics[width=7cm,clip]{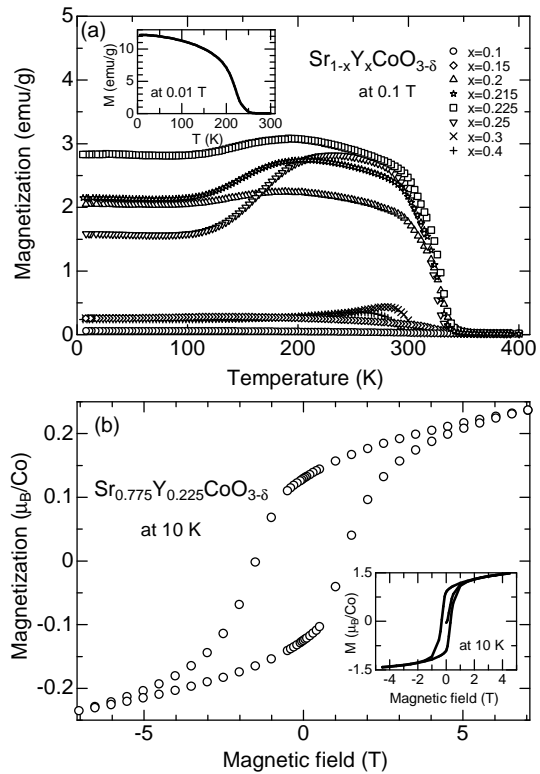}
\caption{(a) Magnetization versus temperature ($M$($T$)) of Sr$_{1-x}$Y$_x$CoO$_{3-\delta}$ 
($x=$ 0.1, 0.15, 0.2, 0.215, 0.225, 0.25, 0.3 and 0.4) , (b) 
magnetization versus magnetic field ($M$($H$)) of Sr$_{0.775}$Y$_{0.225}$CoO$_{3-\delta}$. The insets of 
Fig. 1(a) and (b) show $M$($T$) and $M$($H$) of Sr$_{0.775}$Y$_{0.225}$CoO$_3$, respectively.}
\end{figure}

Figure 1(a) shows the magnetization of Sr$_{1-x}$Y$_x$CoO$_{3-\delta}$ at 0.1 T 
as a function of temperature. 
A tiny trace of a ferromagnetic transition is seen at 300 K for $x=0.1$ and 0.15,
and a weak and broad peak is seen near 300 K for $x=0.3$ and 0.4, 
which is consistent with the literature \cite{withers, troyanchuk2}. 
Surprisingly, the samples only in a narrow compositional range of $0.2\leq x \leq 0.25$ 
show large magnetization. 
The ferromagnetic transition temperature ($T_c$) is 335 K, 
which is, to our knowledge, the highest temperature in perovskite Co oxides. 

The range of $0.2\leq x \leq 0.25$ is quite singular, below which or above which 
the magnetization rapidly fades away. This makes a remarkable contrast with 
La$_{1-x}$Sr$_x$CoO$_3$, where $T_c$ and magnetization 
continuously change with the Sr content. 
Thus, the observed ferromagnetism should not be attributed to the 
double-exchange mechanism. 

The inset of Fig. 1(a) shows $M$($T$) of Sr$_{0.775}$Y$_{0.225}$CoO$_3$ 
annealed in high-pressure oxygen.
We expect no oxygen vacancies ($\delta=0$) for this sample.
Thus the formal valence of Co is more than 3.7, 
and the magnetism seems to be dominated by the double exchange 
interaction between Co$^{4+}$ and Co$^{3+}$.
In fact, the observed ferromagnetic state is quite different
from that in the oxygen-deficient samples:
$T_c$ is remarkably lower, but the magnetization is larger.
These features are essentially similar to $M$($T$) 
of La$_{1-x}$Sr$_x$CoO$_3$ \cite{LaCoO}.
These differences clearly reveal that oxygen deficiency is essential 
to the peculiar room-temperature ferromagnetism.

Figure 1(b) shows the magnetic-field dependence of the magnetization ($M$($H$)) 
of Sr$_{0.775}$Y$_{0.225}$CoO$_{3-\delta}$ at 10 K. 
At 7 T, the magnetization reaches 0.25 $\mu_B$/Co. 
Although this value is somewhat smaller 
than that of La$_{1-x}$Sr$_x$CoO$_3$ (typically more than 1 $\mu_B$/Co), 
it is large enough to be taken as an indication of bulk ferromagnetism,
and is almost equal to that of polycrystalline sample 
of GdBaCo$_2$O$_{5.5}$ at 250 K \cite{MR, troyanchuk2}. 
The inset of Fig. 1(b) shows $M$($H$) of Sr$_{0.775}$Y$_{0.225}$CoO$_3$ 
(the annealed sample) at 10 K 
in which the $M$($H$) curve is quantitatively different from that in 
the non-annealed samples.
These experimental facts further consolidate that the ferromagnetism 
in Sr$_{1-x}$Y$_{x}$CoO$_{3-\delta}$ is 
not caused by the double-exchange mechanism. 

\begin{figure}
\vspace*{0cm}
\includegraphics[width=6.5cm,clip]{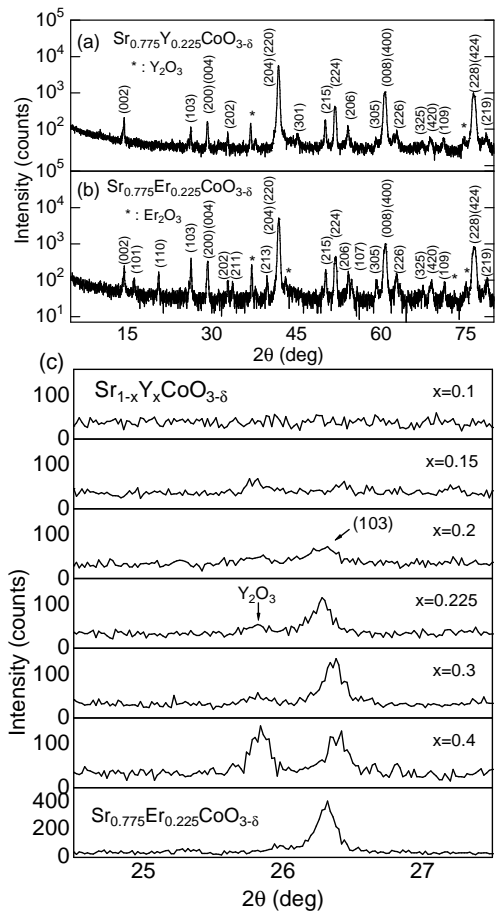}
\caption{ X-ray diffraction patterns of 
(a) Sr$_{0.775}$Y$_{0.225}$CoO$_{3-\delta}$ and (b) Sr$_{0.775}$Er$_{0.225}$CoO$_{3-\delta}$. 
(c) X-ray diffraction patterns around 26$^{\circ}$ of Sr$_{1-x}$Y$_x$CoO$_{3-\delta}$ 
($x=$ 0.1, 0.15, 0.2, 0.225, 0.3 and 0.4) and Sr$_{0.775}$Er$_{0.225}$CoO$_{3-\delta}$.}
\end{figure}

Figure 2(a) shows the XRD pattern of Sr$_{0.775}$Y$_{0.225}$CoO$_{3-\delta}$,
which is consistent with those of Ref. \cite{withers,james,istomin1,istomin2}. 
All the peaks are indexed as a tetragonal cell of the space group I4/mmm 
with the lattice parameters of $a=7.674$ \AA~and $c=15.34$ \AA, 
except for small peaks of Y$_2$O$_3$ (at most 2\%; note that the intensity scale is logarithmic). 
Note that the ($h$00), (0$k$0), (00$l$) ($h, k, l$  $\neq$ 2$n$) peaks do not 
appear according to the extinction rule. 

Figure 2(b) shows the XRD pattern of Sr$_{0.775}$Er$_{0.225}$CoO$_{3-\delta}$, 
where all the peaks except for small peaks of Er$_2$O$_3$ 
(at most 4\%) are again successfully indexed as the tetragonal cell.
The only difference from Fig. 2(a) is that the (101), (110) and (103) peaks 
are much stronger in Fig. 2(b). 
This should be associated with the different X-ray scattering cross sections 
of Er and Y, directly indicating that these 
superstructure peaks come from the A-site ordering. 
In other words, if the A site cations were disordered 
(but the Co or O ions were ordered), 
the two XRD patterns should be identical. 
Although the possibility of the ordering of 
the Co and O ions is not excluded,
the above results strongly support that the A-site ordering is predominant
\cite{comment}.

\begin{table}
\caption{Crystallographic and iodometric titration data for 
Sr$_{1-x}$Y$_x$CoO$_{3-\delta}$ ($x=$ 0.1, 0.15, 0.2, 0.225, 0.25, 0.3 and 0.4). 
Standard deviation is given in parentheses.}
\begin{center}
  \begin{tabular}{l c c c c c}
$x$&a (\AA)&c (\AA)&Space group&3-$\delta$&Co valence\\  \hline
0.1&3.853(1)&-&Pm3m&2.695(3)&3.289\\ 
0.15&3.847(1)&-&Pm3m&2.690(1)&3.230\\ 
0.2&7.685(7)&15.37(1)&I4/mmm&2.655(0)&3.110\\ 
0.225&7.674(8)&15.34(1)&I4/mmm&2.650(3)&3.075\\ 
0.25&7.664(8)&15.32(1)&I4/mmm&2.641(0)&3.033\\ 
0.3&7.647(7)&15.26(1)&I4/mmm&2.643(2)&2.985\\ 
0.4&7.649(7)&15.24(1)&I4/mmm&2.634(2)&2.865\\ 

\end{tabular}
\end{center}
\label{ionradi}
\end{table}%

Figure 2(c) shows the $x$ dependence of the (103) peak, 
which is a direct probe for the A-site ordering.
Clearly, it appears above $x=0.2$, which exactly corresponds to the
emergence of the room-temperature ferromagnetism in this compound. 
As listed in Table I, the crystal symmetry for $x=0.1$ and 0.15 is cubic (Pm3m)
without any superstructure peaks like the (103) peak in Fig. 2(a),
which is consistent with Ref. \cite{istomin2}. 
Thus we conclude that the A-site ordering drives 
the ferromagnetism in the Co ions.
This is the same situation as in the case of $R$BaCo$_2$O$_{5.5}$, 
where the ordering of $R$ and Ba stabilizes the Co$^{3+}$ ordering.
Thus we think that the ferromagnetism in the present material 
would be driven by the orbital ordering, although the ordering pattern
may be different because $R$BaCo$_2$O$_{5.5}$ shows a different XRD pattern.
We further note that the peak is monotonically shifted to a higher degree
with increasing $x$, indicating that the lattice parameter is systematically decreased
by substitution of Y for Sr (See Table I). 
On first glance, it looks inconsistent that the $x=0.3$ sample does not show the bulk 
ferromagnetism in spite of having the same intensity of the (103) peak 
as the $x=0.25$ sample.
We think that the ferromagnetism is quite susceptible 
to the Sr/Y solid solution in the Sr site:
Y cannot help occupying the Sr site for $x>$1/4.
In fact, our conjecture is verified by the fact that
Sr$_{0.75-y}$Ca$_y$Y$_{0.25}$CoO$_{3-\delta}$ rapidly 
loses ferromagnetism with $y$ \cite{kobayashi}.
In contrast, the oxygen content is unlikely to play an important role, 
because it remains essentially constant ($3-\delta=2.64$) between $x=0.25$ and 0.3.

\begin{figure}
\vspace*{0cm}
\includegraphics[width=6.5cm,clip]{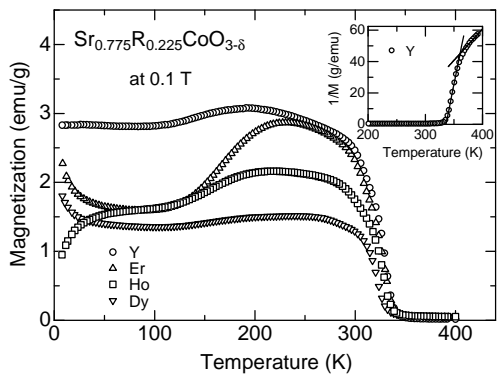}
\caption{ Magnetization versus temperature of 
Sr$_{0.775}R_{0.225}$CoO$_{3-\delta}$ ($R$ = Dy, Ho, and Er). 
The inset shows 1/$M$ of Sr$_{0.775}$Y$_{0.225}$CoO$_{3-\delta}$.}
\end{figure}

Figure 3 shows the $M$($T$) of Sr$_{0.775}R_{0.225}$CoO$_{3-\delta}$ 
($R$ = Dy, Ho and Er). Not only the Y-substituted samples but also the other compounds 
exhibit a similar ferromagnetism below 335 K. 
This means that the magnetic moments of the lanthanides are 
irrelevant to the 335 K transition. 
The inset of Fig. 3 shows 1/$M$ of Sr$_{0.775}$Y$_{0.225}$CoO$_{3-\delta}$,
in which a kink is observed at 360 K. 
A similar kink was observed 
in 1/$M$ of TbBaCo$_2$O$_{5.5}$ at the orbital-ordering temperature 
\cite{LnBaCoO1}. 

The low-temperature magnetization is affected by the magnetic 
moment of Dy$^{3+}$, Ho$^{3+}$, and Er$^{3+}$. 
The low-temperature susceptibility of the Er- and Dy-substituted 
samples obeys the Curie law below about 50 K.
The effective magnetic moment
is evaluated to be 6.7 $\mu _B$/Er for the Er-substituted sample, 
corresponding to 70 \% of that expected from Er$^{3+}$ of $J=15/2$ and $g=6/5$, 
where $J$ is the total angular momentum and $g$ is Lande's g factor. 
The effective magnetic moment is 5.3 $\mu _B$/Dy for the Dy-substituted sample, 
corresponding to 50 \% of that expected
from Dy$^{3+}$ ($J=$15/2 and $g=$4/3). 
Since the orbital angular momentum is sometimes quenched in the 
presence of ligand field, the magnetic moments 
are reasonably evaluated for the above samples.
On the other hand, the Ho-substituted sample shows 
anomalous susceptibility that decreases with decreasing temperature 
below 50 K.
It may indicate an antiferromagnetic coupling 
between Co$^{3+}$ and  Ho$^{3+}$ ions, and 
such a 3$d$-4$f$ coupling is actually observed in $R_2$Cu$_2$O$_5$
\cite{garcia}.

Figure 4 shows the resistivity of Sr$_{1-x}$Y$_x$CoO$_{3-\delta}$. 
The temperature dependence of resistivity is nonmetallic, 
which is again incompatible with the double-exchange mechanism. 
For $0.2\leq x \leq 0.25$ samples, a resistivity jump is seen at 360 K, 
which is close to $T_c$. 
A similar but larger jump was reported in the resistivity of GdBaCo$_2$O$_{5.5}$
\cite{troyanchuk2, kopcewicz, taskin}. 
We further note that the resistivity for 0.2 $\leq x \leq$ 0.25 is 
clearly distinguished from that for $x\le$ 0.15 or $x\ge$ 0.3. 
The inset of Fig. 4 shows 
the resistivity of Sr$_{0.775}$Y$_{0.225}$CoO$_3$ annealed in 
the high-pressure oxygen. 
The resistivity shows metallic conduction with a kink at 250 K, 
which is very different from the resistivity of the non-annealed samples.

Now we discuss the origin of the room-temperature ferromagnetism in 
Sr$_{1-x}$Y$_x$CoO$_{3-\delta}$ ($0.2\leq x \leq 0.25$). 
As listed in Table I, the oxygen content 3-$\delta$ for $x=$ 0.25 
is $\sim$ 2.64, corresponding to the formal Co valence of $\sim$3.03. 
This is in between the oxygen content of 2.69 for $x=$0.2 
by James et al. \cite{james} 
and the oxygen content of 2.62 for $x=$0.3 
by Istomin et al. \cite{istomin1}.
Very recently, Maignan {\it et al.} \cite{maignan} found
a similar oxygen content of 2.66 for $x$=0.33,
and reported that the formal valence of Co is 2.99.
From these data we believe that the Co ions are essentially trivalent for $x=$ 0.25.
Thus the ferromagnetism most likely arises from the orbital ordering 
of the IS Co$^{3+}$ ions, as in the case of $R$BaCo$_2$O$_{5.5}$,
which naturally explains the several similarities 
to the ferromagnetism in $R$BaCo$_2$O$_{5.5}$ mentioned above.

\begin{figure}
\vspace*{0cm}
\includegraphics[width=6.5cm,clip]{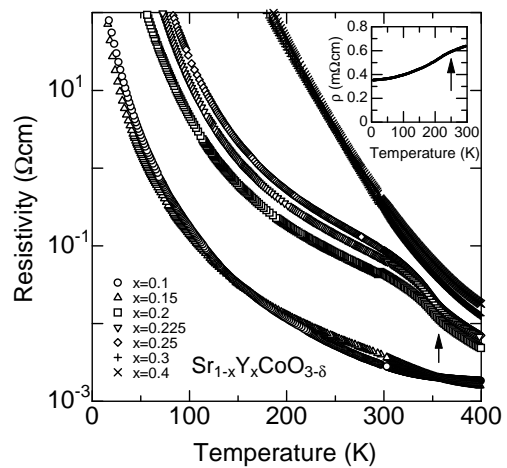}
\caption{The resistivity of Sr$_{1-x}$Y$_x$CoO$_{3-\delta}$ 
($x=$ 0.1, 0.15, 0.2, 0.225, 0.25, 0.3, and 0.4).
The arrow indicates the kink temperature in the inset of Fig. 3.
The inset shows the resistivity of Sr$_{0.775}$Y$_{0.225}$CoO$_3$
The arrow indicates the Curie temperature.}
\end{figure}

Next, we will compare Sr$_{1-x}$Y$_x$CoO$_{3-\delta}$ 
with $R$BaCo$_2$O$_{5.5}$. 
The ferromagnetic order is highly robust down to 5 K 
in Sr$_{1-x}$Y$_x$CoO$_{3-\delta}$, and is most likely the ground state of this system.
According to the theory of orbital ordering \cite{KCuF},
an antiferro-type orbital order favors a ferromagnetic order,
and a ferro-type orbital order favors an antiferromagnetic order.
Suppose the present compound be an insulator consisting of IS Co$^{3+}$.
Then an orbital order will be  antiferro-type,
and prevent formation of the $e_g$ band, which drives a ferromagnetic order. 
Similar ferromagnetic states are seen in other orbital-ordered systems,
such as K$_2$CuF$_4$ \cite{KCuF} and YTiO$_3$ \cite{YTiO}.
In this context $R$BaCo$_2$O$_{5.5}$ is rather exceptional, 
in which the antiferromagnetic order is the ground state 
in spite of the anitiferro-type orbital ordering.
Thus the ferromagnetic state competes with the antiferromagnetic state
in $R$BaCo$_2$O$_{5.5}$, which causes large magnetoresistance.
Such a competition can explain why $T_c$ is lower than $T_c$
of Sr$_{1-x}$Y$_x$CoO$_{3-\delta}$.

Finally, we would like to add a few notes. 
(i) As a preliminary measurement, we measured 
the magnetoresistance of Sr$_{0.775}$Y$_{0.225}$CoO$_{2.65}$ at 300 K. 
The value is $-0.5$ \% at 7 T, which is very small compared with 
that of a typical giant magnetoresistance material. 
This suggests that the ground state is ferromagnetic and insulating,
which is consistent with our orbital-ordering scenario.
(ii) We examined an impurity effect, and observed that 
the ferromagnetism is quite susceptible against impurities, where
only a 1\%-substitution of Mn for Co destroys a half of the saturation magnetization. 
Such result cannot be understood from simple dilution effects, 
and could be attributed to the impurity effect 
on the orbital ordering. In a case of charge ordering, a tiny disorder strongly 
suppresses the charge ordering state \cite{motome}. 

\section{Summary}

We have measured magnetic susceptibility and resistivity 
of Sr$_{1-x}$Y$_x$CoO$_{3-\delta}$ ($x=$ 0, 0.1, 0.15, 0.2, 0.225, 0.25, 0.3 and 0.4), 
and have found that Sr$_{1-x}$Y$_x$CoO$_{3-\delta}$ 
for 0.2 $\leq x\leq$ 0.25 can be regarded as 
a room-temperature ferromagnet. 
The ferromagnetic transition temperature is approximately 335 K, 
which is the highest transition temperature among perovskite Co oxides. 
The magnetization of the $x=$ 0.225 sample at 10 K is 0.25 $\mu_B$/Co, 
which implies a bulk ferromagnet. 
We conclude that the ferromagnetic order is driven by the peculiar Y/Sr ordering.
We expect that this material will open a new window to 
materials science of the A-site ordered Co oxides.

\begin{acknowledgments}

We thank B. Raveau and A. Maignan for fruitful discussion.
This work was partially supported by Grant-in-Aid for JSPS Fellows.

\end{acknowledgments}

\end{document}